**How Marius Was Right and Galileo Was Wrong Even Though Galileo Was Right and Marius Was Wrong -- or, how telescopic observations in the early 17th century supported the Tychonic geocentric theory and how Simon Marius realized this.**

Christopher M. Graney

Jefferson Community & Technical College

1000 Community College Drive

Louisville, KY 40272 (USA)

christopher.graney@kctcs.edu

www.jefferson.kctcs.edu/faculty/graney

ABSTRACT

Astronomers in the early 17th century misunderstood the images of stars that they saw in their telescopes.  For this reason, the data a skilled observer of that time acquired via telescopic observation of the heavens appeared to support a geocentric Tychonic (or semi-Tychonic) world system, and not a heliocentric Copernican world system.  Galileo Galilei made steps in the direction of letting observations lead him towards a Tychonic or semi-Tychonic world system.  However, he ultimately backed the Copernican system, against the data he had on hand.  By contrast, the German astronomer Simon Marius understood that data acquired by telescopic observation supported a Tychonic world system.

The development of the astronomical telescope in the early 17th century precipitated a number of discoveries which undermined the Ptolemaic world system that had dominated astronomy for centuries -- the sun has spots, the moon has mountains, Venus has phases, Jupiter has moons.  These observations made it clear that the heavens are not perfect, the heavens are not unchanging, the heavens are not absolutely unearthly, Venus circles the sun, a miniature system exists around Jupiter.  But while such observations undermined the Ptolemaic system and its Aristotelian foundation, they undermined neither of the most prominent candidates to replace the old system -- the heliocentric Copernican and the geocentric Tychonic world systems.  These two were observationally equivalent regarding the sun, moon, and planets.  Only observations of the stars, such as the successful detection of annual stellar parallax, could distinguish between these two systems.  This paper will show that close observation of the stars in the early 17th century would yield data that strongly supported the Tychonic system, not the Copernican system.  It will argue that Galileo made such observations of stars, and came to some intermediate conclusions leading in the direction of the Tychonic system, even though he ended up supporting the Copernican system.  This paper will argue that the German astronomer Simon Marius[1] also made such observations of stars, and saw that the data supported the Tychonic system.

---

1   Special thanks to Thony Christie who introduced me to Simon Marius's observations of the stars and who shared with me both his copy of *Mundus Iovialis/Die Welt des Jupiter* and his insights into Marius's work.

Previous work from this author posted to arxiv.org or published in *Baltic Astronomy* has put forth a number of arguments concerning telescopic observations of the stars in the early 17th century. Focusing on the work of Galileo Galileo, this work has emphasized three main points. Briefly:[2]

I) Galileo used a small aperture (roughly 30 mm), optically perfect telescope to observe stars. He found that the telescope resolved[3] the stars into distinct disks. Seen through such a

---

[2] **I**, **II**, and **III** summarize work from the author's papers listed below, with the first paper generally giving details of **I**, the second, **II**, etc. The reader can find detailed references, explanatory diagrams, etc. in these:

Christopher M. Graney, "On the Accuracy of Galileo's Observations", *Baltic Astronomy*, vol. 16 (2007), pp. 443-449. Also arXiv:0802.1095 [http://arxiv.org/abs/0802.1095].

Christopher M. Graney, "Visible Stars as Apparent Observational Evidence in Favor of the Copernican Principle in the Earth 17th Century", *Baltic Astronomy* vol. 17 (2008), pp. 425-438. Also arXiv:0901.2656 [http://arxiv.org/abs/0901.2656].

Christopher M. Graney and Henry Sipes, "Regarding the Potential Impact of Double Star Observations on Conceptions of the Universe of Stars in the Early 17th Century", arXiv:0812.3833 [http://arxiv.org/abs/0812.3833]. The issue of parallax is also dealt with in Christopher M. Graney, "But Still, It Moves: Tides, Stellar Parallax, and Galileo's Commitment to the Copernican Theory", *Physics in Perspective* vol. 10 (2008), pp. 258-268.

[3] In modern terms no early 17th century telescope could *resolve* the disks of stars. However, Galileo pointed the telescope at a star, looked through it, and saw a measurable, magnifyable disk. From the point of

telescope, these disks vary in size with magnitude; brighter stars show larger disks, fainter stars show smaller disks. Such disks respond to magnification, like the larger disks exhibited by the planets. Galileo, being ignorant of the wave nature of light and the phenomenon of diffraction, interpreted these disks as being the physical bodies of stars. Such disks are spurious, having nothing to do with the actual size of the stars. They are manifestations of the Airy disk formed by the telescope. All stars have the same Airy disk diameter, which is a function only of wavelength and telescope aperture. However, because the eye can detect only a limited level of intensity, the spurious disks of brighter stars appear larger than those of fainter stars. Galileo's measurements of these disks (the apparent sizes of stars as seen through his telescope) appear to be accurate -- consistent with what would be expected for a diffraction limited telescope of the size he used. Galileo observed stars to have diameters in the range of a few arc seconds, eventually concluding that stars of magnitude *M = 1* had apparent diameter *α = 5"* in diameter, while stars of *M = 6* had *α = 5/6"*. A linear relationship between *M* and *α* is consistent with what would be expected for a diffraction limited telescope of the size he used, allowing for measurement error. In addition, Galileo observed double and multiple star systems. These observations included Mizar and the Trapezium; he made highly accurate measurements of the positions of stars in these two systems. Galileo could measure stellar positions and apparent sizes to an accuracy of at least 2".

II) Galileo hypothesized that all stars are suns scattered throughout space (all essentially the same size as the sun).

---

view of a 17$^{th}$ century astronomer, the telescope resolved the star's disk just as it resolved the disk of Jupiter. It is from that perspective that the word *resolve* is used in this paper.

Based on this hypothesis, the number of stars $N^*$ within a distance $L$ of Earth would be expected to increase as $L^3$. Galileo thought he was seeing the physical bodies of stars with his telescope, and so thought stars' apparent sizes relative to the apparent size of the sun indicated their distances: the distance $L$ (in AU) of a star with apparent telescopic diameter $\alpha$ is given by $L = \alpha_\odot/\alpha$, where $\alpha_\odot$ is the apparent diameter of the sun.[4]  $N^*$ goes as $L^3$, $L$ goes as $1/\alpha$, and $\alpha$ decreases linearly with $M$ (as mentioned in **I**), so therefore

$$N^* \sim \frac{1}{(C-M)^3} \qquad \text{(equation 1)}$$

where $C$ is a constant.  This equation, and the hypothesis of stars being suns that underpins it, can be tested by counting visible (naked-eye) stars.  The results confirm Galileo's ideas about the stars -- equation 1 matches the actual increase in number of visible stars with magnitude to within the uncertainty brought on by error in measuring stellar diameters.[5] Additionally, the telescope reveals stars not visible to the unaided eye, suggesting that the stars extend indefinitely, even infinitely, into space, so the idea that Earth could be at rest with the stars rotating about it diurnally becomes absurd.  Thus telescopic observation provides evidence that the Earth has motion -- a diurnal rotation upon its own axis.

---

4   Note that Galileo did not use the inverse square law to determine stellar distances, just simple geometry.

5   This "confirmation" is a matter of coincidence as Galileo's ideas are based on the erroneous view that his telescope can resolve the physical bodies of stars.

III) The distances Galileo determined for the stars using $L = \alpha_\odot/\alpha$ are large (hundreds to thousands of AU), but not large enough to explain the lack of annual stellar parallax.  As mentioned in **I**, Galileo's observations of the stars included a small number of stars that are double, with a smaller star in close proximity to a larger star.  Such a pairing might be interpreted as two stars of different sizes that are close in all dimensions of space, but as the telescopic data suggests that the stars are essentially identical suns scattered through space, such a pairing must be a line-of-sight alignment of a nearer and farther star.  However, at the distances implied by $L = \alpha_\odot/\alpha$, the stars in such a line-of-sight system should exhibit parallax, and the difference between the stars' parallaxes (differential parallax) should greatly exceed the stars' separations.  That no differential parallax is detected implies that the Earth is not moving relative to the fixed stars.  Thus while the data indicates that the Earth has a diurnal rotation, that data also indicates that the Earth is not circling the sun.

Therefore, according to the data provided by the telescope, the early 17th century astronomer must consider that the true world system is as follows:

- ✔ The Earth is fixed in location but rotates diurnally.
- ✔ The moon circles Earth monthly.
- ✔ The sun circles Earth yearly, with its Tychonic retinue of moons (planets), some of which have their own moons in turn (e.g Jupiter).
- ✔ The stars are suns, scattered throughout space.
- ✔ The universe is indefinitely large, at least thousands of AU in diameter -- possibly infinite.

This is essentially the Tychonic system, but with a rotating Earth and an indefinitely large universe of stars.  This semi-Tychonic system is the world system towards which observational data would drive any early 17th century astronomer who undertakes a sufficiently thorough program of telescopic observations -- so were any astronomers thus driven?

Galileo made the telescopic observations.  He saw and measured the stars' spurious disks.  He interpreted those disks as being the physical bodies of the stars.  He hypothesized that stars were suns scattered through space.  He used that hypothesis to determine stellar distances.  He observed double stars.  He saw a relationship between a star's apparent size and its magnitude.[6]  But he went no further.  He did not follow his observations to the logical conclusions.  He backed Copernicus, relying on his unconvincing tidal theory to serve as evidence of Earth's motion.  He argued that the stellar distances he calculated supported Copernicus.[7]  He even suggested that observations of double stars could yield convincing evidence of

---

6   That Galileo had such ideas may surprise the reader -- modern discussions of Galileo's major astronomical work tend either to overlook Galileo's views and discoveries concerning the stars, or to limit themselves to his description of them in the *Starry Messenger* (1610), his first publication of his telescopic observations.  A large collection of quotes from Galileo's writings concerning the stars can be found in Graney, "Visible Stars...".

7   See Galileo Galilei, *Dialogue Concerning the Two Chief World Systems*, translated by S. Drake (University of California Press, Los Angeles, 1967), p. 359, and Galileo Galilei, "Rely to Ingoli" in *The Galileo Affair:  A Documentary History,* by M. A. Finocchiaro (University of California Press, Los Angeles, 1989), pp. 166-168.

Earth's motion[8], when in fact he had observations of double stars in his notes that gave witness in just the opposite direction[9].

Simon Marius of Anspach, Germany also made the telescopic observations.  Unlike Galileo, Marius followed those observations to the logical conclusion that telescopic observations supported a Tychonic world system.

Literature about Marius's work is scarce.  What exists typically notes that he observed the Jovian system, that the names used for Jupiter's "Galilean" moons originated with him, that he recorded seeing the Andromeda galaxy, and that he incurred Galileo's ire.  A hunt for analysis of Marius's work turns up a few references that suggest his skill at telescopic observation rivaled Galileo's:  Marius observed the spurious disks of stars[10]; from his observations of the Jovian system he derived better periods for and certain elements of the orbits of the Jovian moons than did Galileo[11].

Marius's comments about stellar observations contained in his *The Jovian World* (1614) are brief.  They are not well known -- the commonly available English translation of *The*

---

8    Galileo, *Dialogue...* p. 382.

9    As discussed previously in **I** and **III**.

10   J. L. E. Dreyer, "The Tercentenary of the Telescope", *Nature* (December 16, 1909), p. 191.

11   A. Pannekoek, *A History of Astronomy* (Interscience Publishers, New York, 1961), p. 231.

*Jovian World* does not include them.[12]  Marius writes that gaining possession of a good telescope made it possible for him to see something which he could not see earlier -- that the telescope resolves as disks not only the planets, but also the brighter stars.  He goes on to write that he is truly surprised that Galileo has not seen this, noting that Galileo writes in the *Starry Messenger* that the stars do not possess a defined circular shape.  Marius also notes that this is supposedly part of the strongest argument in favor of the Copernican world system -- the lack of a round shape to the stars indicates that they lie at an immense distance from Earth, as Copernicus says they must.  But, says Marius, the fact that the disks of the stars can be seen from Earth undercuts this argument, and bolsters its reverse -- the fixed stars are clearly not at the immense distances required by Copernicus, and instead their appearance agrees with the Tychonic world system.  Marius notes additional evidence for the Tychonic system in the moons of Jupiter, whose motions he finds consistent with their circling Jupiter while the Jupiter circles the sun.  Marius adds that these ideas will require further discussion and explanation.  Lastly, he concedes to Galileo that the stars shine by their own light -- they are distinct in appearance from the planets (being notably more intense in brilliance).[13]

---

12  "The 'Mundus Jovialis" of Simon Marius", translated by A. O. Prickard, *The Observatory* Vol. 39 (1916), pp. 367-381, 403-412, 443-452, 498-503 [http://adsabs.harvard.edu/abs/19160bs....39..367].

13  This paragraph summarizes Marius's original text from *Mundus Iovialis*, which is as follows:

> *Tertium est, quod non ita pridem, videlicet post reditum à*

Evidently, like Galileo, Marius observed the spurious disks of stars with a telescope. Like Galileo, he interpreted the disks as being the physical bodies of stars. Like Galileo, he observed that disk size varies with magnitude.[14] And like Galileo, he concludes that stellar disks reveal something about stellar distances. Marius did not leave the quantity of

---

> *Ratisbona mihi pararim instrumentum, quo non solum planetae, sedetiam, omnes fixae insigniores exquisitae rotundae cernuntur, inprimis autem canis major, minor, lucidiores in Orione, Leone, Ursamajore, etc. quod antehac nunquam mihi videre contigit. Miror equidem Galilaeum cum suo instrumento admedum excellente idem non vidisse. Scribit enim in suo Sidereo Nuncio, fixas stellas periphaeria circulari nequaquam terminatas apparere, id quod quidam posteae maximi argumenti loco habuerunt, nimirum hoc ipso systema mundanum Copernicanum confirmari, nempe quod ob immensam distantiam fixarum à terrâ, figura globosa fixarum stellarum nequaquam in terris ullo modo percipi possit. Cum vero nunc certissime constet, etiam fixas orbiculari in terris hoc perspicillo videri, cadit profectò haec argumentatio, et plane contrarium astruitur, nimirum sphaeram stellarum fixarum nequaquam adeo incredibili distantiâ à terris removeri, uti fert speculatio Copernici, sed potius talum esse segregationem sphaerae fixarum à terris, ut nihilominus moles corporum illarum hoc instrumento figura circulari distinctè videri possit, consentiente ordinatione sphaeraru coelestium, Tychonica et propriâ, ut inferius parte secunda hujus libelli, phaenomeno quinto confirmabitur* [this references the part of the book where Marius discusses how the motions of the Jovian moons support the Tychonic system]. *Verum haec alibi disputanda et explicanda sunt. Quod autem fixae proprio luceant lumine, Galilaeo facile concessero, quia longe excellentiore splendore et claritate sunt praeditae, quam planetae.*

Original Latin provided in the German translation by J. Schlor, *Die Welt*

commentary on stellar observations that Galileo left, but these brief comments indicate that Marius saw what Galileo saw.

However, Marius comes to a conclusion that Galileo does not.  Marius concludes that the appearance of the stars undermines, not supports, the Copernican world system.  Marius uses the stars to argue for a Tychonic system, on the basis that the telescopic appearance of stars shows that they are not distant enough to satisfy the requirements of a Copernican universe.

Unfortunately Marius does not give the details of his argument in this regard, but we can surmise a few things from what he does give us.  That stars show disks is not in itself a convincing argument for anything.  The stars could be very large -- either large bodies at various great distances[15], or large lights of varying size on a stellar sphere.  This does not particularly bolster either the Copernican or the Tychonic world systems.  But Marius states that because the stars show disks they cannot be distant enough to satisfy the requirements of a Copernican universe.  As size alone cannot indicate distance,

---

    *des Jupiter* (Schenk-Verlag, Gunzenhausen, 1988) pp. 46-48.

14   As he notes that these disks are most prominent in the brighter stars, it is reasonable to conclude this.

15   There was precedent for such an idea -- in 1576 Thomas Digges had argued that the stars were bodies far larger than the sun, scattered through space.  See F. R. Johnson, S. V. Larkey, and T. Digges, "Thomas Digges, the Copernican System, and the Idea of the Infinity of the Universe in 1576", *The Huntington Library Bulletin* No. 5 (April 1934), pp. 69-117 [http://www.jstor.org/stable/3818095].

and as Marius was a careful observer who looked for consistency between theory and observation (as can be seen in his study of the Jovian moons), we may conclude[16] that Marius has followed the observations through to at least some of the conclusions reached here.  Marius's views are much more akin to what the data demands than are Galileo's.

    In conclusion, we see that, absent an understanding of the wave nature of light and the Airy disk, early telescopic observers who possessed both high-quality telescopes and sufficient observing skill should conclude that they can resolve

---

16    Marius does not specify that the stars are fixed and the Earth rotates -- he merely references the Tychonic view.  It is possible that Marius does not view stars as being suns, but rather merely self-luminous bodies. We should also note that equation 1 is not actually predicated on the stars being suns.  It only requires that the stars be identical in size.  Once lack of differential parallax has convinced us that the Earth is not moving, we are free to bring the stars as close as we wish -- so long as they are identical and their distances and sizes are proportionally unchanged.  Stars could be, not suns at hundreds to thousands of AU distant, but smaller luminous bodies at tens to hundreds of AU distant.  This may be the preferable arrangement, for if the stars are suns we must ask why one circles Earth and the rest to do not (they will not be circling their own Earths either -- at distances of a few hundred to a few thousand AU, such motion would be noticeable in even a telescope of the early 17$^{th}$ century).  Whether such a condensed universe revolves about Earth or whether Earth rotates within it is a matter of taste.  Which is more distasteful, the idea of ourselves moving at substantial speed (1000+ mph at the equator), or the idea of the furthest stars moving at untold multiples of that speed?  At any rate, if the universe is Tychonic, such a thing could never be observationally determined from Earth's surface.

the bodies of stars.  Furthermore, once early telescopic astronomers accept that they can resolve the bodies of stars, careful observations both of how the numbers of naked-eye stars increase with magnitude and of the lack of differential parallax in close doubles should drive them towards certain conclusions about the universe.  Those conclusions are that the arrangement of the sun, moon, and planets are geocentric in a Tychonic or semi-Tychonic fashion, while the stars are suns scattered throughout space.  We see further that Galileo made the observations needed to reach these conclusions.  However, he supported Copernicus in opposition to the data his telescope provided him.  On the other hand, it seems Simon Marius also made such observations.  Unlike Galileo, Marius followed the data to its Tychonic conclusion.  Marius may well have been every bit Galileo's equal in observing skill, and essentially his superior in terms of maintaining a clear-eyed, data-driven approach to learning about the world.

It is clear that in the early 17$^{th}$ century, thoughtful, thorough, logical acquisition and analysis of telescopic data could lead to the wrong conclusions (and in the case of Simon Marius, very likely did lead to the wrong conclusions).  The work presented here suggests that in the early 17$^{th}$ century the geocentric Tychonic system may have had significant scientific appeal over the heliocentric Copernican system.[17]  At the dawn of

---

17  Decades after Galileo's death the geocentric theory, in Tychonic or semi-Tychonic forms, seems to still have had significant staying power, despite Galileo's famous support of the Copernican world system.  Examples of the Tychonic system being treated with respect well after Galileo's death can be found in J. B. Riccioli's *Almagestum Novum* (1651), telescopic astronomy, for any astronomer with the determination to engage in a thorough study of the stars, and to follow the data wherever it led, the geocentric Tychonic theory, not the heliocentric Copernican theory, was the theory that observations supported.

---

R. Hooke's *An attempt to prove the motion of the earth from observations* (1674), and J. G. Doppelmayr's *Atlas Coelestis* (1742).